\documentclass[
amsmath,amssymb,
aps,
prd, 
twocolumn, 
amsmath,  
nofootinbib,
superscriptaddress
]{revtex4-2}

\usepackage[utf8]{inputenc}
\usepackage{graphicx}% Include figure files
\usepackage[T1]{fontenc}
\usepackage{float}
\usepackage{color}
\usepackage{aas_macros} 
\usepackage[normalem]{ulem}
\usepackage{bm}
\usepackage{float}

\usepackage{hyperref}% add hypertext capabilities
\hypersetup{colorlinks=true, citecolor=blue}

\newcommand{\pr}[1]{\ensuremath{\left[#1\right]}}
\newcommand{\pc}[1]{\ensuremath{\left(#1\right)}}
\newcommand{\chav}[1]{\ensuremath{\left\{#1\right\}}}

\DeclareMathOperator{\E}{\mathbb{E}}
\DeclareMathOperator{\DKL}{D_{\text{KL}}}

\begin{document}

\title{
Conditional variational autoencoder inference of neutron star equation of state from astrophysical observations
}

\author{Márcio Ferreira}
\email{marcio.ferreira@uc.pt}
\affiliation{CFisUC, 
	Department of Physics, University of Coimbra, P-3004 - 516  Coimbra, Portugal}

\author{Micha{\l} Bejger}
\email{bejger@camk.edu.pl}
\affiliation{INFN Sezione di Ferrara, Via Saragat 1, 44122 Ferrara, Italy}
\affiliation{Nicolaus Copernicus Astronomical Center, Polish Academy of Sciences, Bartycka 18, 00-716, Warsaw, Poland}
 
\date{\today}

\begin{abstract}
We present a new inference framework for neutron star astrophysics based on conditional variational autoencoders. Once trained, the generator block of the model reconstructs the neutron star equation of state from a given set of mass-radius observations. While the pressure of dense matter is the focus of the present study, the proposed model is flexible enough to accommodate the reconstructing of any other quantity related to dense matter equation of state. Our results show robust reconstructing performance of the model, allowing to make instantaneous inference from any given observation set. 
\end{abstract}

%\keywords{Suggested keywords}%Use showkeys class option if keyword
                              %display desired
\maketitle

%\tableofcontents

\section{\label{sec:introduction} Introduction}
Neutron stars (NSs) remain one of the most intriguing astrophysical objects whose properties are still far from being understood. One of the many open questions yet to be answered is their composition. The equation of state (EoS) is the key quantity that defines the properties of dense and asymmetric nuclear matter realized inside NSs. The search for the true EoS, among the different candidates, is mainly restricted, at moderate and high baryonic densities, by observations of massive NSs, such as PSR~J1614-2230 with  $1.908 \pm 0.016 M_{\odot}$ \cite{Demorest2010,Fonseca2016,Arzoumanian2017}, PSR~J0348 - 0432 with $2.01 \pm 0.04M_{\odot}$ \cite{Antoniadis2013}, PSR~J0740+6620 with $2.08 \pm0.07 M_{\odot}$ \cite{Fonseca:2021wxt}, and J1810+1714 with $2.13 \pm0.04M_{\odot}$ \cite{Romani:2021xmb}. The observation of gravitational waves (GWs) signals produced during the last stages of binary NSs systems by LIGO/Virgo collaboration, such as GW170817 \cite{Abbott:2018wiz} and GW190425 \cite{Abbott:2020khf}, supplied additional constrained on the EoS. It is expected that future GW observations and future experiments, e.g., enhanced X-ray Timing and Polarimetry mission (eXTP) \cite{eXTP,eXTP:2018anb}, the STROBE-X \cite{STROBE-X}, and Square Kilometer Array \citep{SKA} telescope,  might contribute with additional constraints on the EoS due to higher-precision NS observations. \\

Bayesian inference \cite{RevModPhys.83.943} is often used to estimate the entire probability distribution over the EoS parameter space given a set of  NS observational and/or theoretical constraints. The specified prior distributions over the EoS parameters are updated with the Bayes theorem using observed new data. The resulting EoS posterior reflects the updated beliefs about the EoS parameters given the data, see, e.g., \cite{Steiner:2010fz,Malik:2022zol,Malik:2023mnx}. The posterior distributions are usually approximated by Markov Chain Monte Carlo (MCMC) methods, which are the standard technique for Bayesian inference for sampling from probability distributions. However, these methods are computationally prohibitive for large data sets and large parameter sets (high dimension space of parameters), requiring effort to reach convergence and accurate estimates.\\

Several advances in machine learning (ML) techniques have led to impressive developments in different areas during the last decade, e.g., autonomous driving, image recognition, language models. In addition to industrial applications, the potential of ML in different fields of science has definitely captured the interest of researchers, namely in high-energy physics \cite{Zhou:2023pti} and GW astrophysics \cite{Cuoco2021MLSTreview}. 
In NS physics, the use of Gaussian processes as agnostic representations has been explored to represent the EoS of NS matter subjected to several constraints \cite{Essick2019,Landry:2020vaw}, These kernel methods, however, show some specific limitations, e.g., 
the representational power of the kernel function and the substantial computational cost for large datasets.
Several studies explore the application of neural networks (NNs) as inference tools for NS properties and nuclear matter properties from NS observations \cite{Ferreira:2019bny,Fujimoto:2017cdo,Fujimoto:2019hxv,Fujimoto:2021zas,Soma:2022qnv,Morawski:2020izm,Krastev:2021reh,Ferreira:2022nwh,Krastev:2023fnh,Han:2021kjx,Soma:2022vbb,MorawskiB2022PhRvC,Thete:2023aej}.
The same task has been investigated with bayesian neural networks (BNN) that is capable of associating an uncertainty to its predictions \cite{Carvalho:2023ele,Carvalho:2024kgf}. 
Generative ML models, like the autoencoders (AE) or variational autoencoders (VAE) were 
adopted in the past to NS and EoS problems, e.g. to nonparametric representations of NS EoS \cite{2023ApJ...950...77H} and parametrizing postmerger signals from binary NS \cite{2022PhRvD.105l4021W}. 
A framework capable of inferring the EOS 
directly from telescope observations, based on normalizing flows model coupled with Hamiltonian Monte Carlo methods, was introduced in \cite{Brandes:2024vhw}.
Here, we implement a conditional VAE (cVAE) to infer values of dense-matter EoS parameters from a set of macroscopic (astrophysical) observations of NSs, such as their masses and radii, which are functionals of the microscopic EoS parameters.\\

The paper is organized as follows. The VAE model and its conditional variants are introduced in Sect.~\ref{sec:VAE}. The dataset generation and the cVAE training procedure are detailed in Sect.~\ref{sec:dataset}. Section \ref{sec:results} discusses the results and the conclusions are drawn in Sect.~\ref{sec:conclusions}.

\section{\label{sec:VAE} Variational autoencoders}

The VAE is a likelihood-based deep latent-variable model that belongs to the family of generative models \cite{kingma2013auto}. Generation and inference is made possible by the introduction of a hidden (unobserved) latent variable $\bm{z}$. For the input data $\bm{x}$, the VAE is composed of two distinct but complementary parts: the probabilistic encoder $q_{\bm{\phi}}(\bm{z}|\bm{x})$ and probabilistic decoder $p_{\bm{\theta}}(\bm{x}|\bm{z})$; see Fig.~\ref{fig:diagram_vae}. The generative process is defined by the decoder $p_{\bm{\theta}}(\bm{x}|\bm{z})$ as it, in the ideal case, generates back (reconstructs) the input data $\bm{x}$ given a sample $\bm{z}$ from the latent space $p(\bm{z})$. Both encoder and decoder are implemented by NNs, where ${\bm{\phi}}$ and ${\bm{\theta}}$ denote the respective NNs weights. In variational inference, one is interested in computing the marginal likelihood posterior distribution (evidence) $p_{\bm{\theta}}(\bm{x})=\int p_{\bm{\theta}}(\bm{x}|\bm{z})p(\bm{z})d\bm{z}$, which however is not tractable due to lack of direct access to the $\bm{z}$ distribution. VAE \cite{kingma2013auto} solves this problem by estimating $p_{\bm{\theta}}(\bm{x})$ through the evidence lower bound (ELBO,  also called the variational lower bound) procedure, as 
\begin{align}
    \log p_{\bm{\theta}}(\bm{x}) &\ge {\cal L}_{\text{VAE}}(\bm{\theta},\bm{\phi},\bm{x}) 
\end{align}
where
\begin{align}
 {\cal L}_{\text{VAE}}(\bm{\theta},\bm{\phi},\bm{x}) :=&\E_{\bm{z}\sim q_{\bm{\phi}}(\bm{z}|\bm{x})}\pr{\log p_{\bm{\theta}}(\bm{x}|\bm{z})} \nonumber \\
 &- 
    \DKL \pc{q_{\bm{\phi}}(\bm{z}|\bm{x})|| p(\bm{z}) }.
\end{align}
The term $\E_{\bm{z}\sim q_{\bm{\phi}}(\bm{z}|\bm{x})}\pr{\log p_{\bm{\theta}}(\bm{x}|\bm{z})}$ is the expectation for the log-likelihood (reconstruction loss), and $\DKL \pc{q_{\bm{\phi}}(\bm{z}|\bm{x})|| p(\bm{z})}$ is the Kullback-Leibler (KL) divergence \cite{10.1214/aoms/1177729694}, which acts a regularizer by forcing the encoder distribution to be as close as possible to $p(\bm{z})$. We use an isotropic multivariate Gaussian distribution for the latent space, $p(\bm{z})={\cal N}(\bm{z};\bm{0},\bm{I})$. 
As shown in \cite{kingma2019introduction}, the use of stochastic gradient descent method \cite{2016arXiv160904747R} is unpractical when optimizing ${\cal L}_{\text{VAE}}(\bm{\theta},\bm{\phi},\bm{x})$, as its gradients with respect to $\bm{\phi}$ show high variance. This can be avoided if the variable $\bm{z}\sim q_{\bm{\phi}}(\bm{z}|\bm{x})$ is re-parameterized by a deterministic and differentiable function $g_{\bm{\phi}}(\bm{x}, \bm{\epsilon}^{(l)})$, where $\bm{\epsilon}^{(l)}\sim {\cal N}(\bm{0}, \bm{I})$. This solution is known as the {\em reparametrization trick}, allowing to easily compute derivatives across the Gaussian latent variables. Consequently, the ELBO can be efficiently obtained with the stochastic gradient variational Bayes estimator \cite{{kingma2013auto}}.

The $\beta-$VAE is a modification of the VAE that adds an adjustable parameter $\beta$ into ${\cal L}_{\text{VAE}}$. This extra hyper-parameter controls the encoding capacity of the latent space and encourages factorized latent representations \cite{burgess2018understanding}. The objective function becomes 
\begin{align}
  {\cal L}^{\beta}_{\text{VAE}}(\bm{\theta},\bm{\phi},\bm{x}, \beta)&=\E_{\bm{z}\sim q_{\bm{\phi}}(\bm{z}|\bm{x})}\pr{\log p_{\bm{\theta}}(\bm{x}|\bm{z})} \nonumber\\
    &-\beta \DKL \pc{q_{\bm{\phi}}(\bm{z}|\bm{x})|| p(\bm{z})}.
\end{align}
This extra term, which forces factorized latent representations, was shown important in different studies. For $\beta>1$, a stronger constrain is set on the disentanglement of the latent space at a cost of a lower reconstruction performance, i.e., $\beta$ introduces a trade-off between reconstruction quality higher  disentanglement representations \cite{burgess2018understanding}. 

\subsection{\label{sec:cVAE} Conditional Variational Autoencoders}

In a cVAE \cite{NIPS2015_8d55a249}, the encoder and decoder are, in addition to being trained on the input data $\bm{x}$, conditioned upon a variable $\bm{y}$, i.e., $q_{\bm{\phi}}(\bm{z}|\bm{x},\bm{y})$ and $q_{\bm{\theta}}(\bm{x}|\bm{z},\bm{y})$, respectively. The loss function remains the same, but on its conditional form:
\begin{align}
    {\cal L}^{\beta}_{\text{cVAE}}(\bm{\theta},\bm{\phi},\bm{x},\bm{y},\beta)&=\E_{\bm{z}\sim q_{\bm{\phi}}(\bm{z}|\bm{x},\bm{y})}\pr{\log p_{\bm{\theta}}(\bm{x}|\bm{z},\bm{y})} \nonumber \\
    &-\beta\DKL\pc{q_{\bm{\phi}}(\bm{z}|\bm{x},\bm{y})|| p(\bm{z})}.
    \label{eq:Loss}
\end{align}

\begin{figure*}[!htb]
    \centering
    \includegraphics[width=1.0\linewidth]{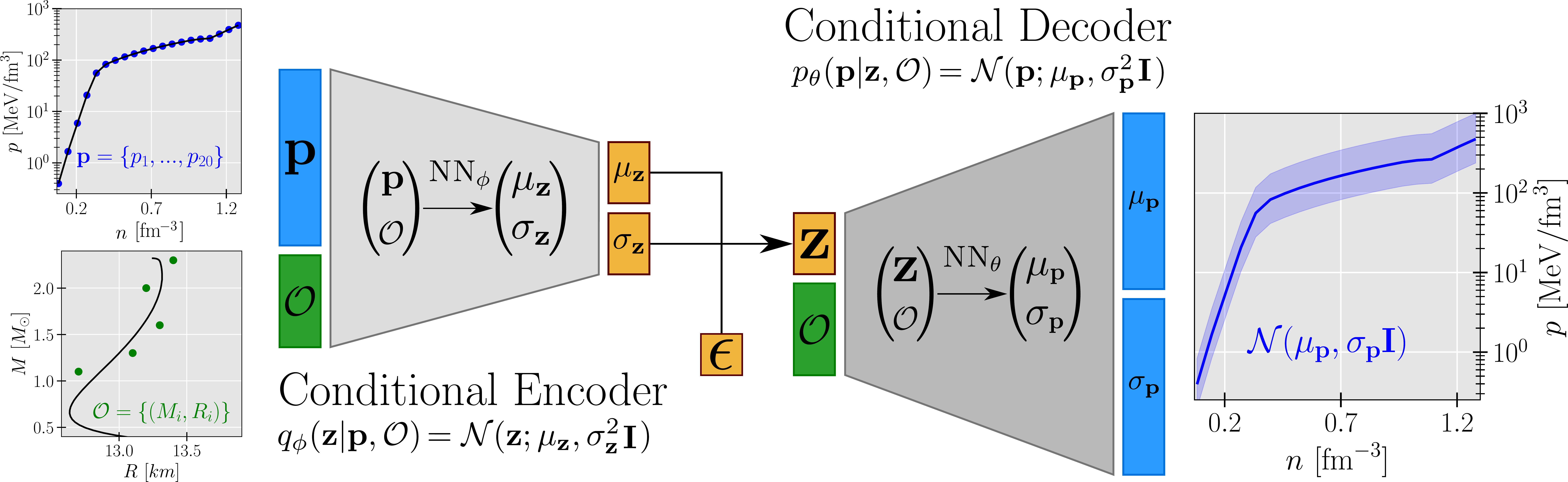}
    \caption{The cVAE model used in this work. The inputs $\bm{x}$ are the EOS pressure $\bm{p}$ and an observation ${\cal O}$, simulating a set of $M(R)$ observations. The output consists of a multivariate Gaussian with variance ${\bm{\sigma_p}}$ and mean ${\bm{\mu_p}}$. See Sect.~\ref{sec:cVAE} for details on the cVAE loss function and training methodology, and Sect.~\ref{sec:EoS} for the training data details.}
    \label{fig:diagram_vae}
\end{figure*}

We will explore this type of model in the present work. To understand exactly how we intend to apply this model in NS physics, we show its structure in Fig.~\ref{fig:diagram_vae}.   
The (probabilistic) encoder and (probabilistic) decoder, which are parameterized by NNs, are conditioned upon a given {\it observation} vector ${\cal O}$, while the input vector is denoted by $\bm{p}$. The encoder is defined as $q_{\bm{\phi}}(\bm{z}|\bm{p},{\cal O}) = {\cal N}\pc{\bm{z};\bm{\mu_z},\bm{\sigma_z}^2\bm{I}}$, where $\bm{\mu_z}$ and $\bm{\sigma_z}$ are outputs of a NN with weights $\bm{\theta}$, i.e., $(\bm{\mu_z},\bm{\sigma_z})=\text{NN}_{\bm{\theta}}(\bm{p},{\cal O})$. There is an external random variable $\bm{\epsilon}\sim {\cal N}(\bm{0},\bm{I})$ and the latent space vectors are given by $\bm{z}=g(\bm{\phi},\bm{\epsilon},\bm{p},{\cal O})=\bm{\mu_z}+ \bm{\sigma_z}\odot \bm{\epsilon} $ (reparametrization trick), where $\odot$ denotes element-wise product.  
Similar to the enconder, the decoder is given by $p_{\bm{\theta}}(\bm{p}|\bm{z},{\cal O}) = {\cal N}\pc{\bm{p};\bm{\mu_z},\bm{\sigma_z}^2\bm{I}}$, where both $\bm{\mu_z}$ and $\bm{\sigma_z}$ are also outputs of a NN with weights $\bm{\phi}$, i.e.,  $(\bm{\mu_p},\bm{\sigma_p})=\text{NN}_{\bm{\phi}}(\bm{z},{\cal O})$. 

Because both $q_{\bm{\phi}}(\bm{z}|\bm{p},{\cal O})$ and $p(\bm{z})$ are Gaussian, the KL divergence term from Eq.~\ref{eq:Loss} can be integrated analytically,   
\begin{align}
\DKL&\pc{q_{\bm{\phi}}(\bm{z}|\bm{p},{\cal O})|| p(\bm{z})} =
\frac{1}{2}\biggr[||\bm{\mu_z}(\bm{p},{\cal O}) ||^2 \nonumber\\
&+||\bm{\sigma_z}(\bm{p},{\cal O}) ||^2 - D - \sum_{k=1}^D\log \pc{\sigma_z^2(\bm{p},{\cal O})}_k\biggr],
\end{align}
where $D$ is the dimension of the vectors $\bm{\mu_z}$ and $\bm{\sigma_z}$, and $||\bm{a}||=a_1^2+\cdot\cdot\cdot+a_D^2$. The objective function in Eq.~\ref{eq:Loss}, for a given data-point $\bm{p}^{i}$ and conditioning vector ${\cal O}^{j}$, takes the following form
\begin{align}
&{\cal L}^{\beta}_{\text{cVAE}}(\bm{\theta},\bm{\phi},\bm{p}^{i},{\cal O}^{j},\beta) = \frac{1}{L}\sum_l^L\log {\cal N}\pc{\bm{p}^{i}|\bm{z}^{ijl},{\cal O}^{j}} \nonumber\\
&- \frac{\beta}{2}\pr{||\bm{\mu_z}^{ij} ||^2 +||\bm{\sigma_z}^{ij} ||^2 - D - \sum_{k=1}^D\log \pr{\pc{\sigma_z^{ij}}^2}_k},
\label{eq:final_loss}
\end{align}
where $\bm{z}^{ijl}= \bm{\mu_z}^{ij}+\bm{\sigma}^{ij}\odot \bm{\epsilon}^{l}$, with $\bm{\epsilon}^{l}= {\cal N}(\bm{0},\bm{I})$, and $L$ is the total number of samples from the latent space. The notation was simplified by $\bm{\mu_z}^{ij}\equiv\bm{\mu_z}(\bm{p}^i,{\cal O}^j)$ and $\bm{\sigma_z}^{ij}\equiv\bm{\sigma_z}(\bm{p}^i,{\cal O}^j)$.
We use $L=1$ during training, which is enough as long as the batch size is large \cite{NIPS2015_8d55a249}. 
Training the model consists in minimizing the objective function Eq.~\ref{eq:final_loss} (loss function) with respect to $\chav{\bm{\theta},\bm{\phi}}$, using stochastic gradient variational Bayes \cite{{kingma2013auto}}.

\section{\label{sec:dataset} Dataset and Training}

\subsection{\label{sec:EoS} Generating the training data}

We have created a training dataset of dense-matter EoS based on piecewise polytropes, $p(\rho)=K\rho^{\Gamma}$, where $\rho=mn$ is the rest-mass density, $n$ baryonic number density, $m$ mass of a baryon, $\Gamma$ the adiabatic index and $K$ politropic pressure coefficient (see \cite{Read:2008iy,Hebeler:2013nza} for details). For a given EoS, $\bm{p}$ is a vector representation of its pressure vs. baryonic density, $p(n)$, and ${\cal O}$ specifies a set of related $M_i(R_i)$ values, solutions of the Tolman-Oppenheimer-Volkoff (TOV) equations \cite{1939PhRv...55..374O,1939PhRv...55..364T} describing spherically symmetric hydrostatic equilibrium of stars in the general relativity. In order to represent the (unknown) EoS of dense matter, we consider the EoS as five connected polytropic segments.

The first polytrope is defined within the range of densities  $[n_{\text{crust}},1.1n_0]$, where $n_0=0.16$ fm$^{-3}$ is the so-called nuclear saturation density, and $n_{\text{crust}}\equiv n_0/2$, with the polytropic index $\Gamma_0$ randomly-chosen from a range $1.0<\Gamma_0<4.5$, so that it lies within the band allowed by the chiral effective field theory (EFT) interactions \cite{Hebeler:2013nza}. For densities lower than $n_{\rm crust}$ we assume the EoS is known, and apply the SLy4 EoS \cite{sly4}.

The remaining four polytropic segments begin at random densities, such that $n_1<n_2<n_3<n_4$, with randomly-chosen polytropic indices $\chav{\Gamma_1,\Gamma_2,\Gamma_3,\Gamma_4}$. In addition to the constraints mentioned above, the EoS must be consistent with the observation of a $1.97M_{\odot}$ NS, and its speed of sound $v_s=\sqrt{\partial p/\partial e}$, where $e$ is the mass-energy density, remains smaller than the speed of light. We covered the entire parameter space $\chav{\Gamma_0,n_1,\Gamma_1,n_2,\Gamma_2,n_3,\Gamma_3,n_4,\Gamma_4 }$ by randomly sampling from uniform distributions consistent with $\chav{n_1,n_2,n_3,n_4}$ being within $n_0$ and $8n_0$, $1.0<\Gamma_0<4.5$, and $0.05<\Gamma_i<8$ for $i=1,...,4$. This parametrization is very flexible and robust, and the inclusion of $\Gamma_i\approx 0$ values allow for EoS models admitting substantial matter softening, i.e. as in EoS featuring phase transitions \cite{Hebeler:2013nza}.

We have generated a total of 39573 valid EoS, which were randomly divided into 90\%/10\% for training/validation set. For each EoS, the datasets consist of the pressure as a function of baryonic density, $p(n)$, and the respective TOV solution, the $M(R)$ diagram. We represent the continuous $p(n)$ functions by their values at $20$ fixed and equally spaced baryonic densities, $n_i=0.081+0.0631(i-1)\,\text{fm}^{-3}$, $i=1,\dots, 20$. Each EoS is then specified by $\bm{p}= [p(n_1),p(n_2),...,p(n_{20})]$. The EoS $p(n)$ relation may be recovered from these 20 points with interpolation with great accuracy. The present framework allows, however, the number of density points to be increased at a cost of training a larger cVAE model.

The model is not trained on the entire $M(R)$ sequence, but on a discrete subset of {\it mock (synthetic) observations} ${\cal O}^j$. The procedure for generating them is as follows: we sample $k$ NS masses, $M_k^{0}/M_{\odot}$, $k=1,\dots, 5$, from a uniform distribution between $1.0$ and $M_{\text{max}}$ (i.e., $M_k^{0}/M_{\odot} \sim \mathcal{U}{[1,M_{\text{max}}]}$); subsequently, we sample $R_k \sim \mathcal{N}\left(R\left(M_k^{0}\right),\sigma_{R}^2 \right)$ and $M_k \sim \mathcal{N}\left(M_k^{0},\sigma_{M}^2 \right)$ for the {\it observed} radii and masses. As a result, a set of observations ${\cal O}^j=\pr{M_1,...,M_5,R_1,...R_5}$ characterizes a given EoS. Five $M(R)$ pairs were chosen to reduce the training time; this number is realistic from a point of view related to the state-of-art of astronomical observatories. 

One instance of mock observations i.e., a set of five $M(R)$ pairs of values deteriorated by observational errors, is clearly insufficient to train an accurate map between the $M(R)$ sequence and the corresponding EoS. Therefore, we generate $n_{\text{s}}$ realizations of observations for each EoS, varying the $M(R)$ values. Each EoS is thus characterized by $\chav{{\cal O}^1,{\cal O}^2,...,{\cal O}^{n_s}}$. We have used $n_{\text{s}}=200$, $\sigma_{R}=0.1$ km and $\sigma_{M}=0.1M_{\odot}$. The standard deviation $\sigma_{R}=0.1$ km is motivated from the expected uncertainties of future observations (e.g., Einstein Telescope \cite{Branchesi:2023mws}, Cosmic Explorer’s \cite{Evans:2021gyd}). This uncertainty value is just an order-of-magnitude
approximation, while the exact value depends on several properties, e.g., the network detectors configuration, neutron star masses, EoS considered \cite{Huxford:2023qne}. 
Considering different observational uncertainties require retraining the cVAE models in new datasets, or changing the above statistical procedure, e.g., turning $\sigma_{R}$ and/or $\sigma_{M}$ into a random variables \cite{Fujimoto:2019hxv}. Furthermore, we are assuming that real observations are Gaussian, which might not be always a good approximation. A more realistic training set should also contain non-Gaussian observations. Additionally, the underlying empirical astrophysical distribution for $M$ should be considered, instead of the uniform distribution we assumed.
These improvements will be explored in a future study. 

To measure the performance of the final selected model, an independent test set with 4398 EoS was also generated with an importance difference: we used $n_s=1$ to simulate the fact that so far only a very limited number of NS is known. Similar procedures were implemented in \cite{Morawski:2020izm,Fujimoto:2021zas,Carvalho:2023ele,Carvalho:2024kgf}.

\subsection{\label{sec:train_cvae} Training the cVAE}

We trained different architectures for the cVAE model, with different number of hidden layers and number of neurons in both decoder and encoder. Furthermore, the activation functions applied to each layer in both NNs are also hyper-parameters of the model. The standard procedure for finding the best model structure and its hyper-parameters is via the cross-validation: the model that gives lowest loss function (Eq.~\ref{eq:final_loss}) on the validation set is considered the best. During training, we used the ADAM optimizer \cite{kingma2014adam} with a learning rate of 0.001. The models were training for 50 epochs, using a mini-batch size of 2048. In each epoch, the model's weights $\pc{\bm{\theta},\bm{\phi}}$ are updated 3479 times.

The best model found is composed of a conditional encoder $q_{\bm{\phi}}(\bm{z}|\bm{p},{\cal O})$ defined by a NN with three layers with $\chav{30,15,6}$ number of neurons and the rectified linear unit (ReLU) activation functions. The last layer of 6 neurons connects, with identity activation functions, with two additional layers of 6 neurons each, specifying the mean $\bm{\mu_z}$ 
and standard deviation $\bm{\sigma_z}$ of the latent variable $\bm{z}$, defined in a 6-dimensional space ($D=6$ in Eq.~\ref{eq:final_loss}), 
$q_{\bm{\phi}}(\bm{z}|\bm{p},{\cal O})={\cal N}\pc
{\bm{z};\bm{\mu}_{\bm{z}},\bm{\sigma}^2_{\bm{z}}\bm{I}}$, see Fig.~\ref{fig:diagram_vae}. 
The conditional decoder $p_{\bm{\theta}}(\bm{p}|\bm{z},{\cal O})$ is also given by a three-layer NN with $\chav{16, 18, 20}$ number of neurons, with ReLU activation functions. Likewise, the last layer of 20 neurons connects (with identity activation functions) with two additional layers of 20 neurons each, describing the mean $\bm{\mu_p}$ and standard deviation $\bm{\sigma_p}$ of the conditional decoder output, i.e., a 20-dimensional Gaussian distribution with no correlation, $p_{\bm{\theta}}(\bm{p}|\bm{z},{\cal O})={\cal N}(\bm{p};\bm{\mu}_{\bm{p}}, \bm{\sigma}^2_{\bm{p}}\bm{I})$.  The model performance is sensitive to the value of $\beta$, which controls the encoding capacity of the latent space. After an extensive range of tests, we found that $\beta=7$ resulted in the model with best results. 

Once we have a trained cVAE model, the decoder can be used as a conditional generative model: the pressure can be reconstructed from a given set of observations ${\cal O}$ and random vector $\bm{z}$ from the latent space.

\section{\label{sec:results} Results}

To illustrate how cVAE models are trained, let us consider one arbitrary EoS with pressure $\bm{p}^{i}$ and one corresponding mock observation   
${\cal O}^{j}$\footnote{The different indices in $\bm{p}^{i}$ and ${\cal O}^{j}$ are to make clear that any given EoS $i$ has 200 distinct mock observations $j$; we used $n_s=200$ in the dataset generation (see Sect \ref{sec:EoS}).}. During training, $\chav{\bm{p}^{i},{\cal O}^{j}}$ is feed into the encoder, which maps it into $(\bm{\mu_z}^{ij},\bm{\sigma_z}^{ij})=\text{NN}_{\bm{\phi}}(\bm{p}^{i},{\cal O}^{j})$. The latent space is then defined by the NN output as
$ {\cal N}(\bm{z};\bm{\mu_z}^{ij},\bm{\Sigma_z}^{ij}\bm{I})$, where $\bm{\Sigma_z}^{ij}=(\bm{\sigma_z}^{ij})^2$. Next, we draw a sample $\bm{\epsilon}^{1}\sim {\cal N}(\bm{0},\bm{I})$\footnote{The sample is a 6-dimensional Gaussian distribution as the latent space $\bm{z}$ of the trained cVAE is also 6-dimensional (see Sect~\ref{sec:train_cvae}).} that specifies the (sampled) latent vector $\bm{z}^{ij1}=\bm{\mu_z}^{ij}+\bm{\sigma_z}^{ij}\odot\bm{\epsilon}^{1}$. Afterwards, the decoder maps both $\bm{z}^{ij1}$ and ${\cal O}^{j}$ into $(\bm{\mu_p}^{ij1},\bm{\sigma_p}^{ij1})=\text{NN}_{\bm{\theta}}(\bm{z}^{ij1},{\cal O}^{j})$. Lastly, the output distribution is given by 
${\cal N}(\bm{p};\bm{\mu}_{\bm{p}}^{ij1},\bm{\Sigma}_{\bm{p}}^{ij1}\bm{I})$.

\begin{figure}[!htb]
    \centering
    \includegraphics[width=0.95\linewidth]{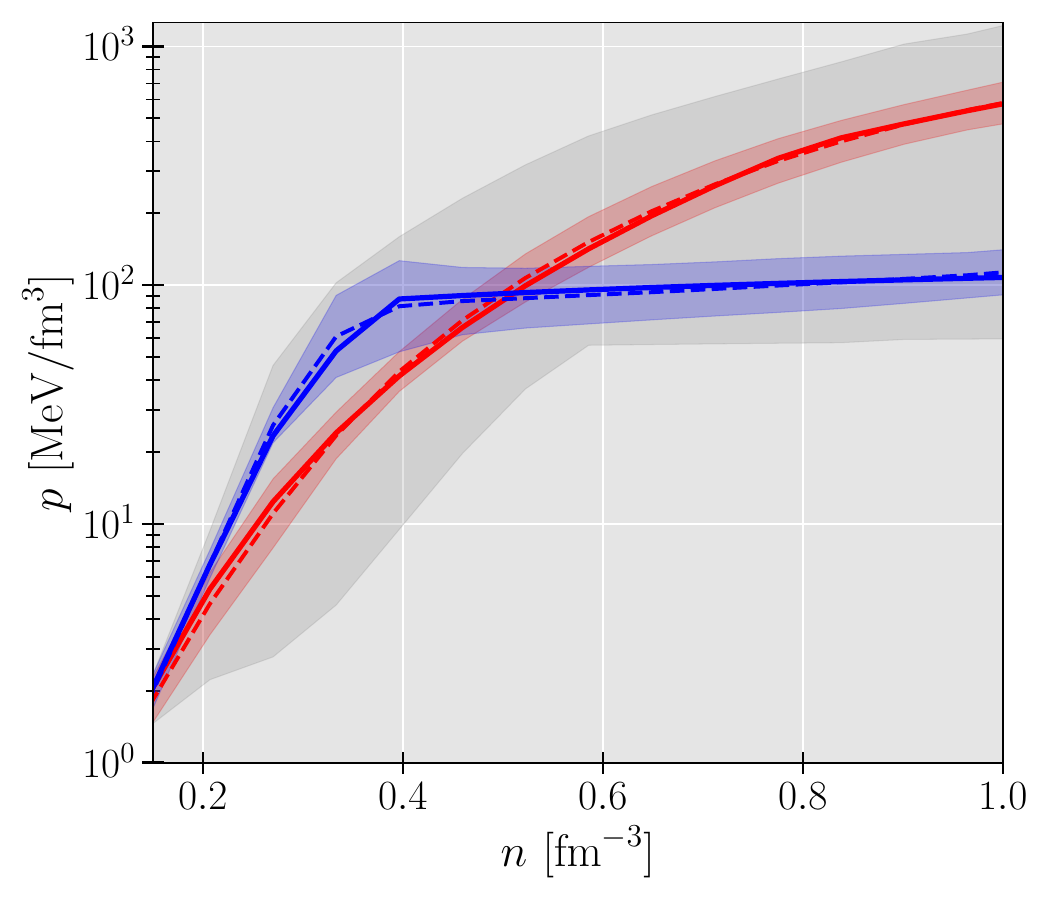}
    \caption{Applying the trained cVAE to reconstruct two EoS from the validation set. The solid lines show the true EoSs (input) and the (Gaussian probability) reconstruction outputs are displayed through its median values (dashed lines) and 95\% CI (shaded region). The gray region represents the range of the training dataset.}
    \label{fig:1}
\end{figure}

A training step consists in updating the weights $\chav{\bm{\phi},\bm{\theta}}$ to minimize the objective function, Eq.~\ref{eq:final_loss}. 
Figure~\ref{fig:1} shows two EoSs (solid curves) that were supplied to the trained cVAE model, as well as the reconstructing output probability distribution, represented by their medians (dashed curves) and 95\% credible intervals (CI\footnote{all credible intervals shown are computed as equal-tailed intervals}).  
Contrarily to the inference phase, discussed next,
the training phase uses a single sample ($L=1$) from the latent space for each $\chav{\bm{p}^i,{\cal O}^j}$.

\begin{figure*}[!htb]
    \centering
    \includegraphics[width=0.32\linewidth]{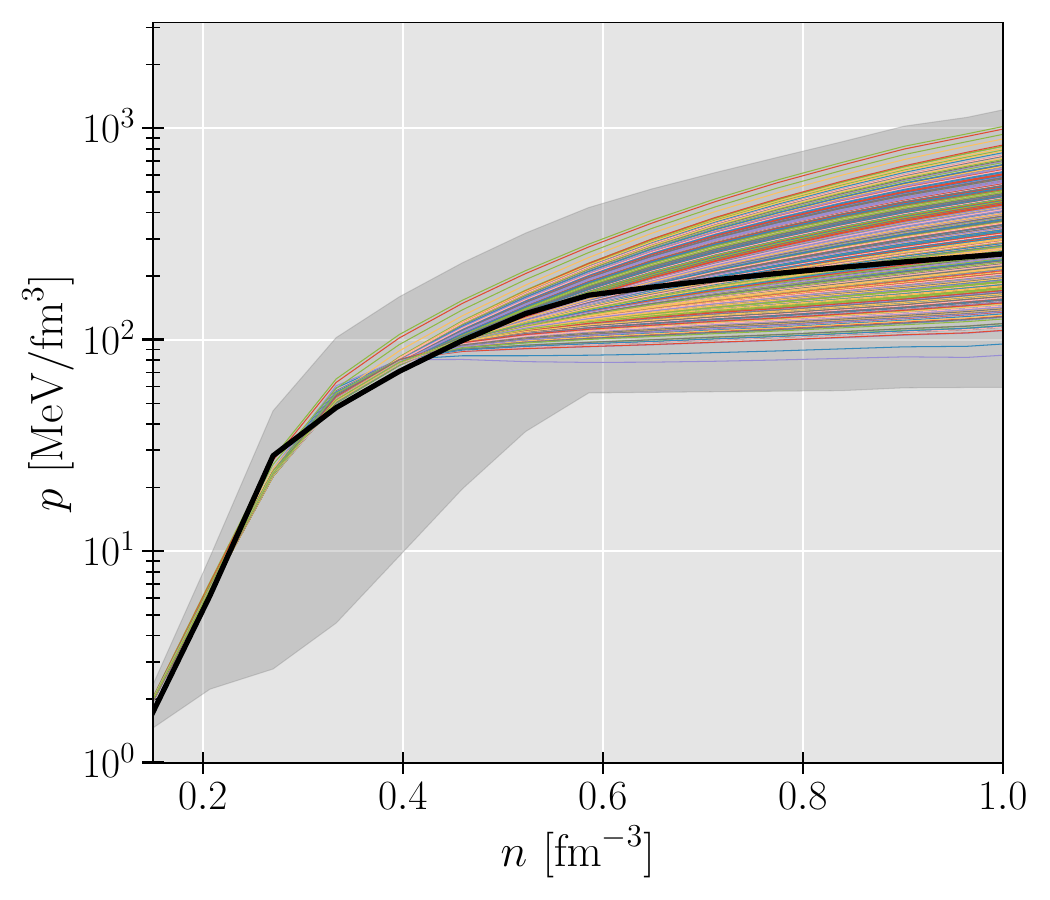}
    \includegraphics[width=0.32\linewidth]{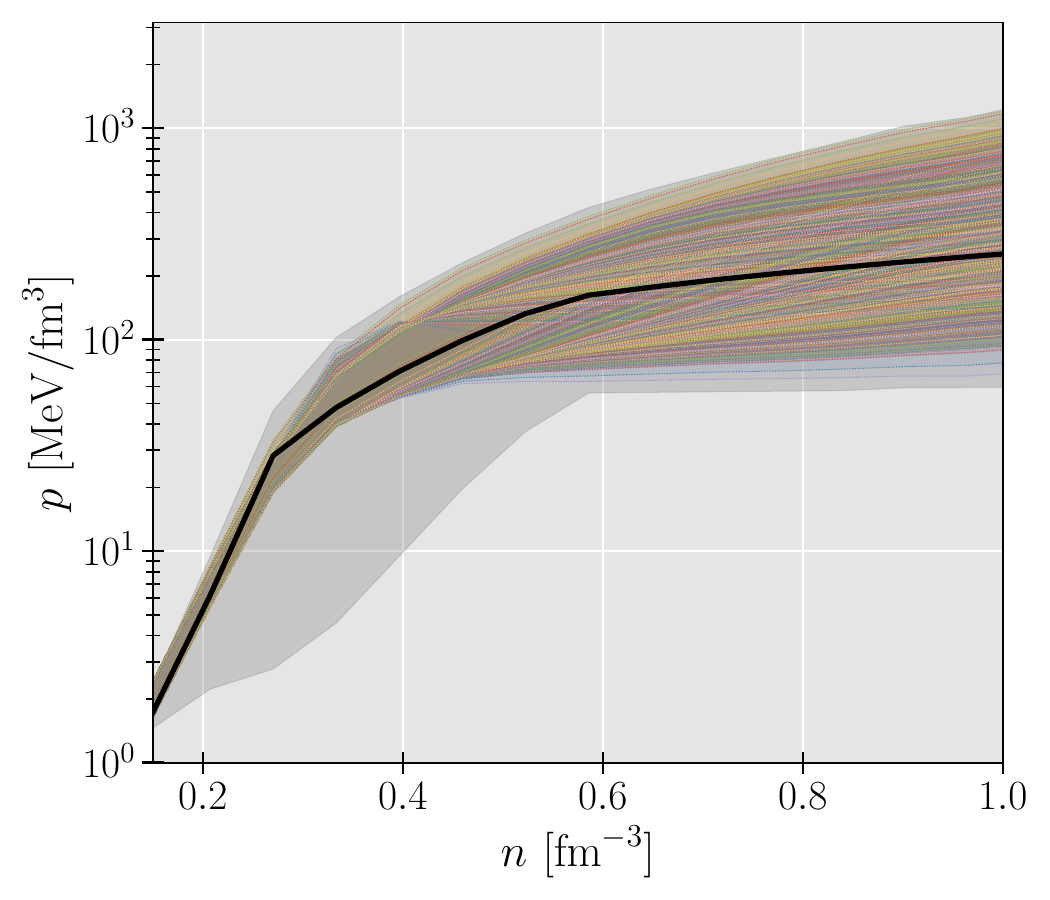}
    \includegraphics[width=0.32\linewidth]{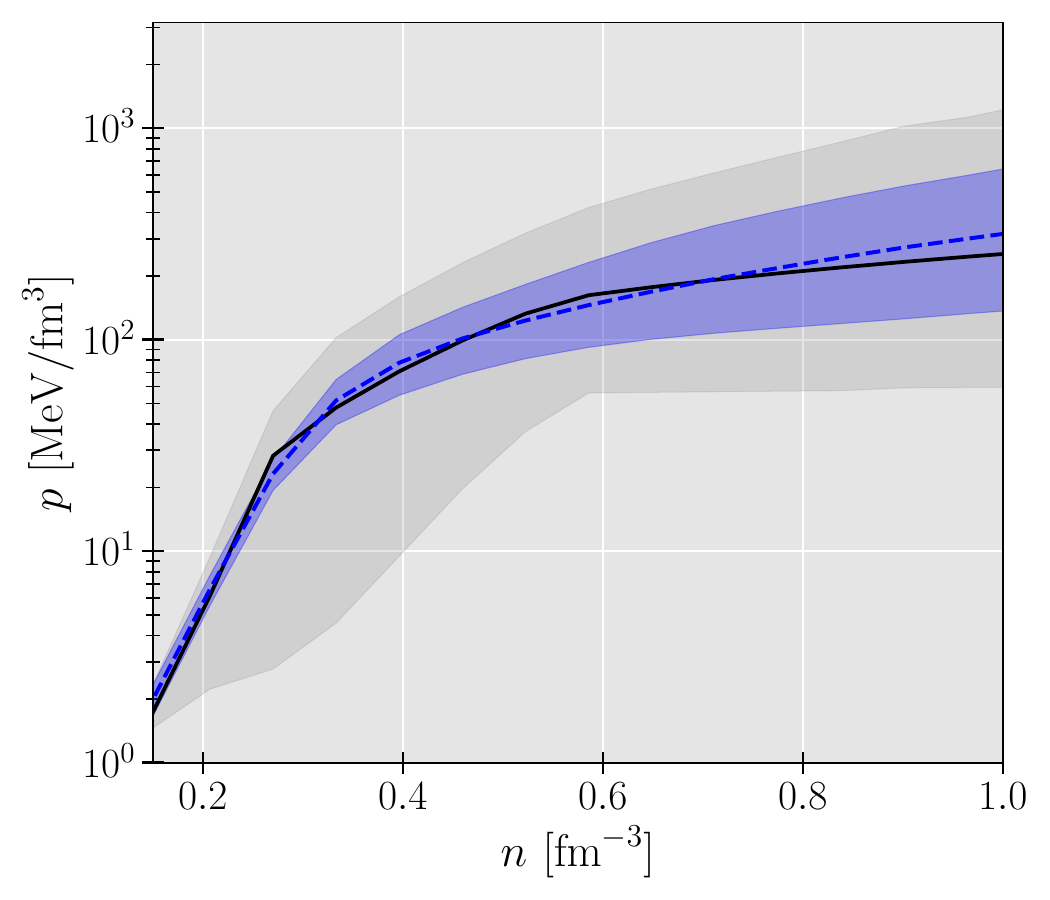}
    \caption{The cDecoder predictive output distribution for one EoS of the test set. 
    We plot the median values (left panel) and 95\% CIs (middle panel) of ${\cal N}_{\bm{\theta}}(\bm{p}|\bm{z}^{l},{\cal O})$, where $l=1,...,2000$. The final reconstruction posterior distribution function,  $p_{\bm{\theta}}(\bm{p}|{\cal O})$, is shown in the right panel. The gray region represents the entire range of the training dataset.}
    \label{fig:2}
\end{figure*}

During the inference phase, where the conditional decoder (cDecoder) is used as generative model, to populate the output space we need more samples from the latent space. 
The predictive distribution is populated by integrating the latent space, $p_{\bm{\theta}}(\bm{p}|{\cal O})=\int {\cal N}_{\bm{\theta}}(\bm{p}|\bm{z},{\cal O}) p(\epsilon)d\epsilon$. We
use Monte Carlo approximation with $L=2000$ samples to estimate it: $p_{\bm{\theta}}(\bm{p}|{\cal O})\approx (1/L)\sum_{l=1}^L {\cal N}_{\bm{\theta}}(\bm{p}|\bm{z}^{l},{\cal O})$, where $\bm{z}^{l}= \bm{\mu_z}+\bm{\sigma}\odot \bm{\epsilon}^{l}$
with $\bm{\epsilon}^{l} \sim {\cal N}(\bm{0},\bm{I})$. The $p_{\bm{\theta}}(\bm{p}|{\cal O})$ consists of a mixture of Gaussian models. Figure \ref{fig:2} shows the reconstruction of an (unknown) EoS (solid black line) from a given observation set ${\cal O}$.
The median values of ${\cal N}_{\bm{\theta}}(\bm{p}|\bm{z}^{l},{\cal O})$ (with $l=1,...,L$) are displayed in the left panel, the the corresponding 95\% CIs in the middle panel, while the final reconstruction probability distribution function, $p_{\bm{\theta}}(\bm{p}|{\cal O})$, 
median and 95\% CI are shown in the right panel. The trained cDecoder was able to 
reconstruct the true EoS from a single set of mock observations ${\cal O}$. 
While the spread of the median values (left panel) at high densities might raise some concern, one should note, however, that the model is performing inference from a single set of mock observations  ${\cal O}$. A stronger constraint is achieved on the high density behavior of the true EoS if massive NSs are within the observation set 
${\cal O}$.

\begin{figure}[!htb]
    \centering
    \includegraphics[width=0.98\linewidth]{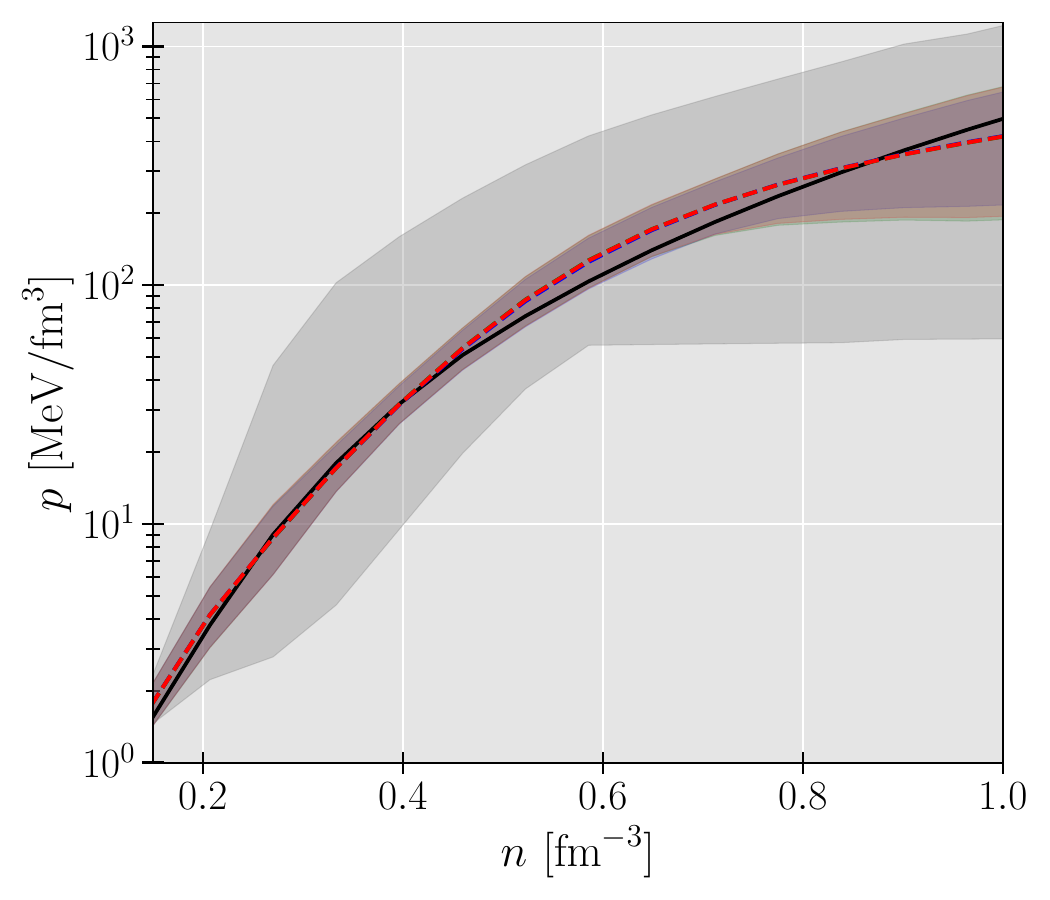}
    \caption{The cDecoder reconstruction output distribution $p_{\bm{\theta}}(\bm{p}|{\cal O})\approx (1/L)\sum_{l=1}^L {\cal N}_{\bm{\theta}}(\bm{p}|\bm{z}^{l},{\cal O})$ with: $L=100$ (blue), $L=1000$ (green), and $L=2000$ (red). We show the reconstructing median values (dashed lines), the 95\% CIs (bands), and the true value (solid line). The gray region represents the range of the training dataset.}
    \label{fig:3}
\end{figure}

Let us analyze the impact of the number of Monte Carlo samples $L$ on the predictive distribution approximation. Figure~\ref{fig:3} shows the inference for a given test set EoS using three values of $L$: 100 (denoted by blue), 1000 (green), and 2000 (black). The deviation between $L=1000$ and $L=2000$ is small, and for values $L>2000$ there is  no significant impact on the the final inference distributions. We set $L=2000$ throughout this work to accurately populate the final probability distribution. 

\begin{figure}[!htb]
    \centering
\includegraphics[width=0.98\linewidth]{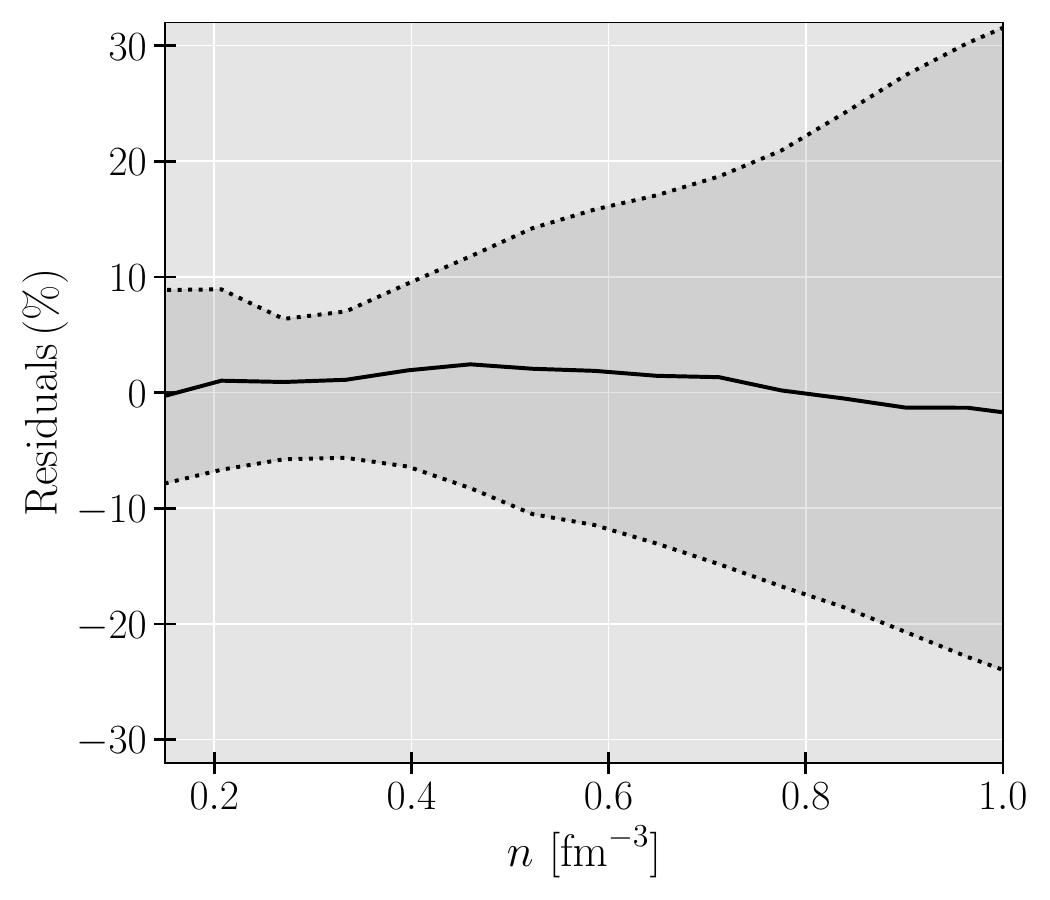}
    \caption{Median (solid lines) and 50\% CI (band) for the residuals of the EoS reconstruction over the test dataset (see text for details). }
    \label{fig:4}
\end{figure}

To analyze the quality of the cDecoder reconstruction, we calculate the residuals ratio (\%), defined by $\bm{\delta}^i=100\times\pc{\mathrm{Med}\pc{p_{\bm{\theta}}(\bm{p}^i|{\cal O})}-\bm{p}^i_{\text{true}}}/\bm{p}^i_{\text{true}}$, for each EoS in the test set. This metric measures the deviation between the predictive distribution median and the true value, in proportion of the true value. Then we determine useful statistics (median and 50\% CI) over the entire test set residuals, $\chav{\bm{\delta}^1,...,\bm{\delta}^N}$, where $N=4398$ is the number of EoS. The results are shown in Fig.~\ref{fig:4}. The residuals median (solid line) is approximately zero for all baryon  densities, and the CI is approximately symmetric around the median value. This shows that, in good approximation, the cDecoder is performing unbiased reconstructions. 
We evaluate the reliability of the predicted CI by determining the number of EOS, from the test set (4398 EOS), confined within the 95\% CI. Considering the entire EOS prediction (all density points), 74\% of the test set is enclosed within the CI, while the fraction increases to 92\% for predictions where at least two density points are outside the CI. The cVAE struggles however to predict abrupt changes in the $p(n)$ from a single observation of five NS.

\begin{figure}[!htb]
    \centering
\includegraphics[width=0.98\linewidth]{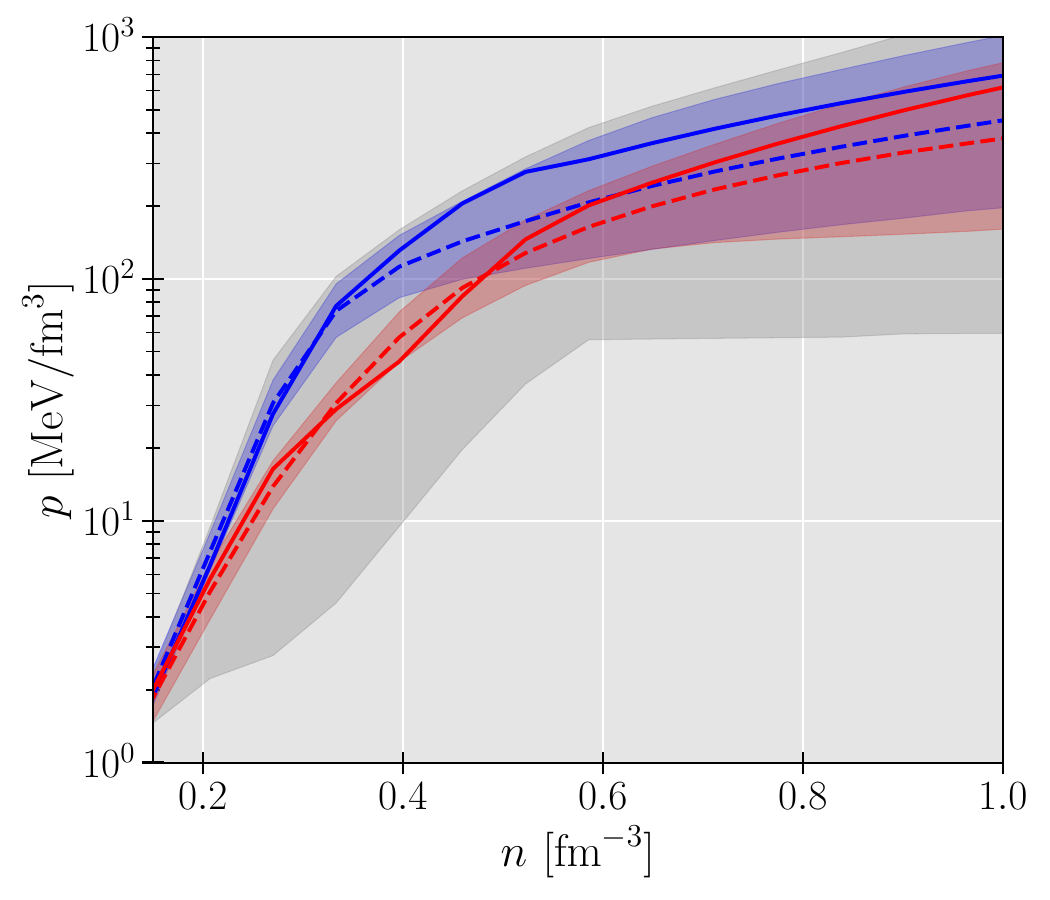}
    \caption{cDecoder applied to two EoS of the test set. The true EoS are displayed in solid curves, the reconstructions medians in dashed curves, and the band displays the 95 \% CI. The gray region represents the range of the training dataset.}
    \label{fig:new}
\end{figure}

To understand the cases where the cVAE predictions are not entirely enclosed within the CI, we show two specific EoS reconstructions in Fig.~\ref{fig:new} where the true EoS lies outside the 95\% CI reconstruction band.
Let us first clarify the difference between Figs.~\ref{fig:new} and \ref{fig:1}. In Fig.~\ref{fig:1} we illustrated how the information flows in the cVAE during training/validating phases: the input data \chav{\bm{p}, {\cal O}} is encoded into the latent space via the cEncoder, then a sample for the latent space, together with ${\cal O}$, enters the cDecoder which predicts the reconstruction EoS. On the other hand, Fig.~\ref{fig:new} shows the inference phase, where the cDecoder reconstructs an EoS from a given observation set. Although the two EoS displayed (solid) were not seen by the cVAE during the training stage, since they belong to the test set, and the inference depends on a single observation ${\cal O}_i$, the true EoSs exhibit only small deviations outside the inferred CI bands. This is a non-trivial result given the flexibility of the polytropic parametrization: due to the randomness of the polytropic indices $\Gamma_i$ at random densities $n_i$, the test set is composed of completely different EoS compared with the training set. Despite that, the cDecoder was able to reconstruct, to a satisfactory extent, these test EoS from a simple set of random 5 NS observations.

\begin{figure*}[!htb]
    \centering
    \includegraphics[width=0.48\linewidth]{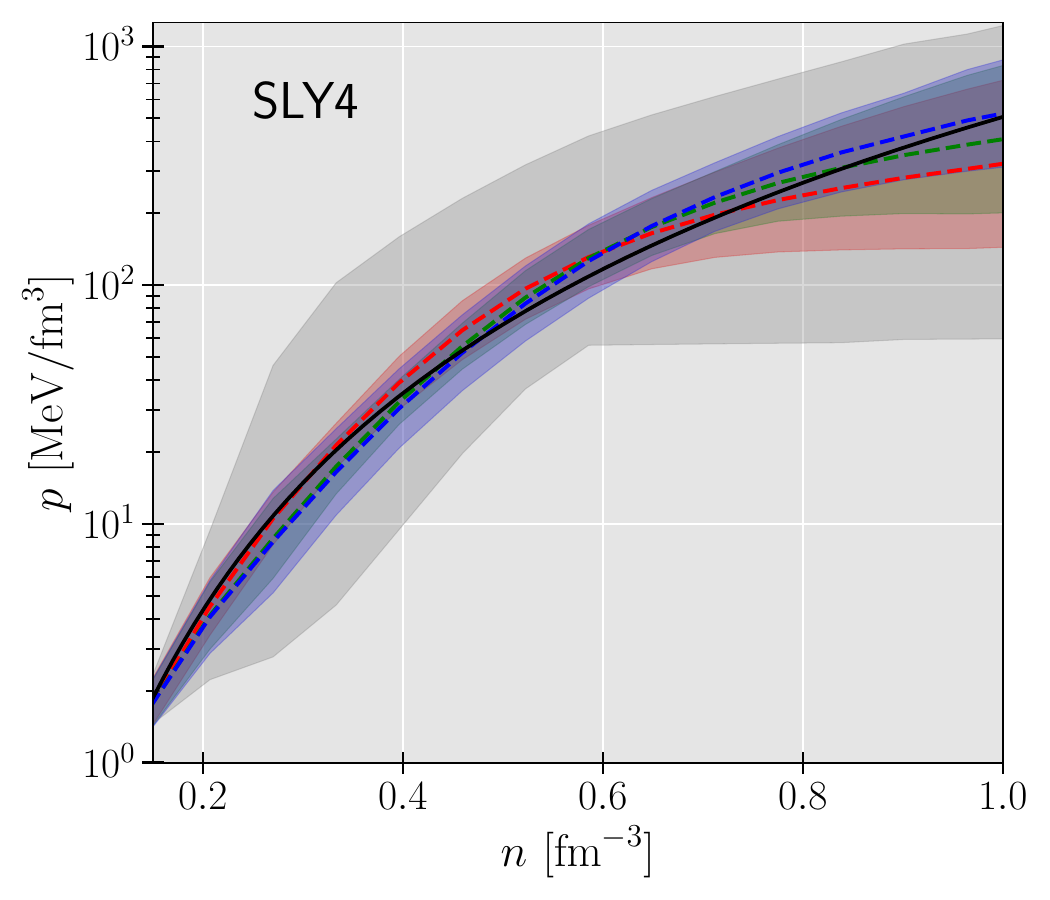}
    \includegraphics[width=0.48\linewidth]{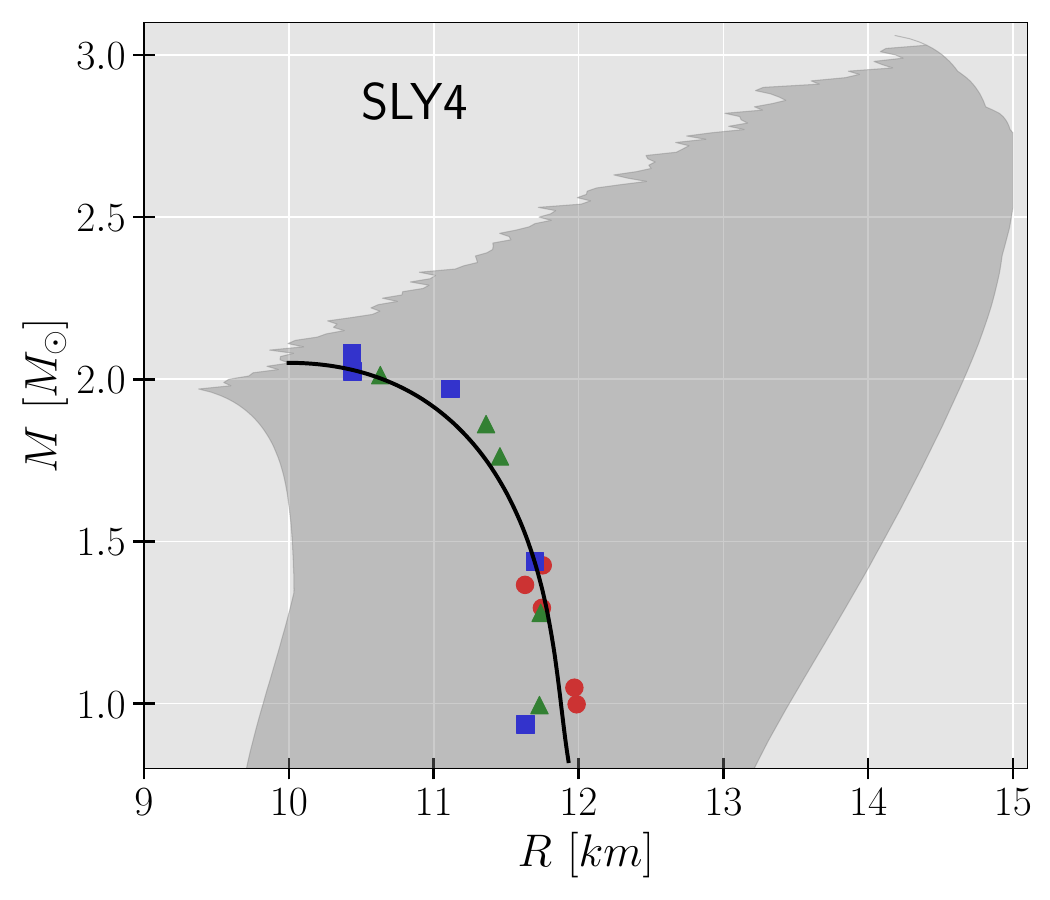}\\
    \includegraphics[width=0.48\linewidth]{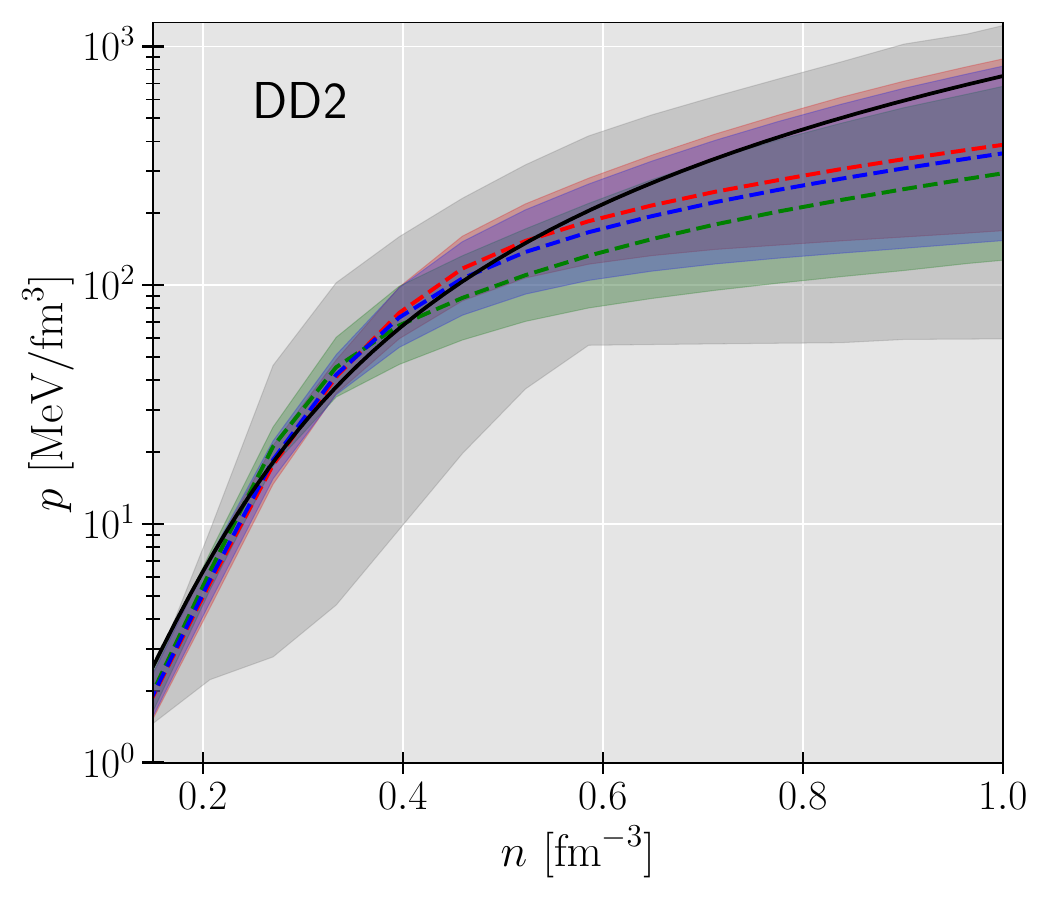}
    \includegraphics[width=0.48\linewidth]{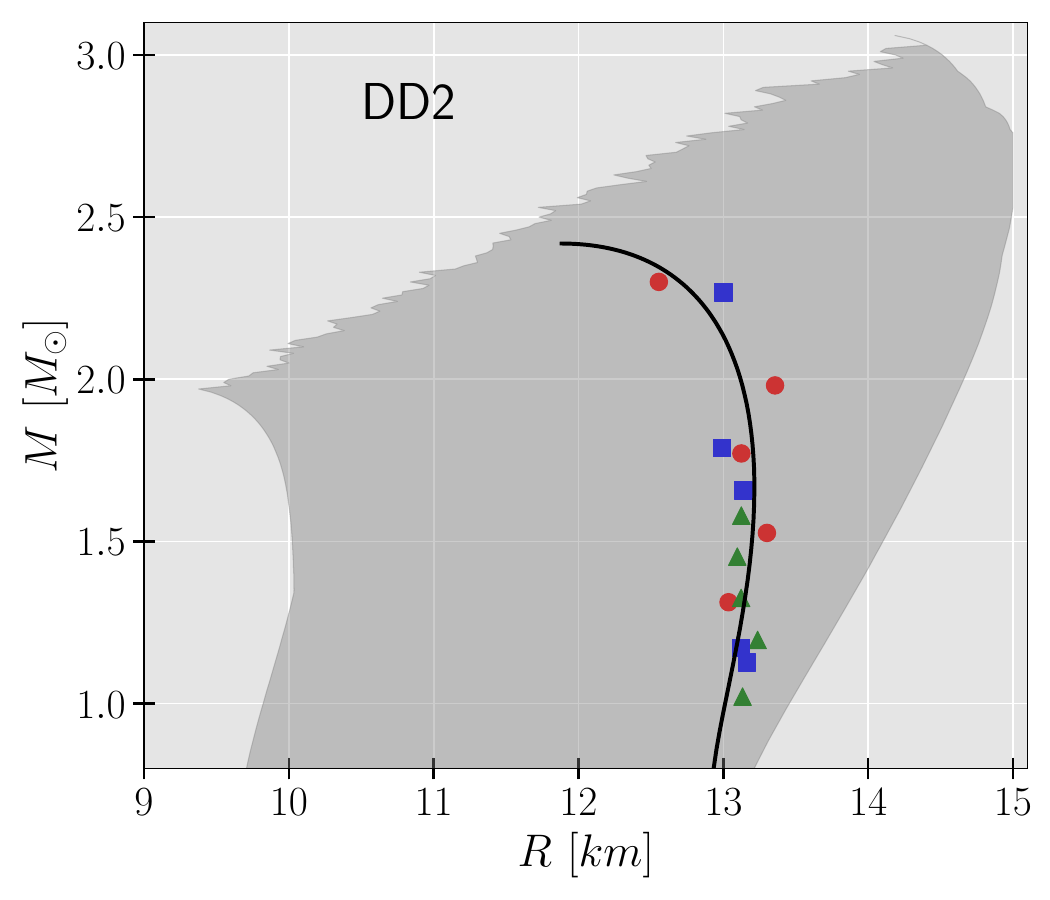}
    \caption{cDecoder reconstruction of the SLy4 and DD2 EoSs (right) from three mock observations ${\cal O}_i$ (right). The colors differentiate the ${\cal O}_i$, for $i=1,2,3$, and respective reconstruction. The gray regions represents the range of the training dataset.}
    \label{fig:5}
\end{figure*}

Finally, we test the cVAE performance in reconstructing two nuclear models EoS. We have selected the DD2 EoS \cite{Typel:2009sy}, a relativistic mean field (RMF) model with density dependent couplings, and the SLy4 EoS \cite{sly4}, a non-relativistic Skyrme interaction model. To illustrate the dependence of the EoS reconstruction upon the mock observation ${\cal O}_i$ set, we have randomly generated three sets. We expect the reconstruction to depend on ${\cal O}_i$: softer EoS, like the SLy4 EoS, are preferred if lower-mass NS are represented, while 
stiffer EoS, like the DD2 EoS, is inferred if massive NS observations are included in the observations. The reconstruction will be independent of ${\cal O}_i$ only if all generated mock observations contain a good snapshot of the whole mass range of the EoS, from low to the maximum NS mass. Figure \ref{fig:5} shows the inferred EoSs (right panels) and the corresponding three mock observations sets $\chav{{\cal O}_1,{\cal O}_2,{\cal O}_3}$ (colours). We note that the cEncoder is able to learn that massive NS observations constrain the high-density dependence of the EoS. Looking at the SLy4 inference, the fact that the red ${\cal O}$ set is composed of NS with $M<1.5M_{\odot}$ translates into a wider reconstruction and slight shifted to lower pressures than the reconstruction for the blue ${\cal O}$ set, which has two NS above $2.0M_{\odot}$. On the other hand, the inference for the red ${\cal O}$ shows smaller deviation at low densities, i.e., the true value is closer to the median prediction, because lighter NS are more informative of this low-density EoS range. The same effect is also present in the DD2 reconstruction, where the 95\% CI encloses the true EoS at high densities when only moderate or massive NS are added to the observation vector {\cal O} (red and blue).

\section{\label{sec:conclusions} Conclusions}
This work describes a proof-of-concept implementation of ML generative models employed for inference models in the domain of NS and dense matter astrophysics. We have applied conditional variational autoencoders for reconstructing the EoS relations from a given set of NS observations, masses and radii. While the present work is focused in inferring the $p(n)$ of NS matter, the extension to other quantities is straightforward.

To train the cVAE models, we have created a dataset of barotropic EoS parameterized by five piecewise polytropes at random densities. This choice was motivated by how generic, robust and capable of reconstructing a large subset of EoS (even phase-transition EoS) the polytropic parametrization is. The EoSs belonging to the training/validation sets are characterized by $n_s=200$ mock observation vectors ${\cal O}_i$, composed of five $M_i(R_i)$ values, which were perturbed by a Gaussian noise to simulate the observation uncertainty. To test the model in a real-case scenario, where only a limited number of NS observations is available, we generated an independent test set of 4398 EoS, each being specified by a single mock observation set ($n_s=1$).

Trained cVAE model has shown great reconstruction performance on the test set, considering that the inference is being conditioned on a single mock observation, i.e.,  five randomly-selected $M_i(R_i)$ values. 
The present framework is quite general and can be easily adapted to predict different NS quantities and include additional information on the conditioning vector, e.g.,  tidal deformability. The model allows to substantially increase the posterior dimension, i.e., the number of density points, at a lower computationally cost. This is an important feature as a fine density grid, and thus a high-dimensional posterior, is required to reconstruct abrupt changes in the EOS. Moreover, once trained, the inference stage of cVAE is instantaneous. 
The developed cVAE-based inference framework is both computationally efficient and capable of scaling with increasing input dimensionality. By compressing high-dimensional data into a concise latent space, the framework reduces complexity while retaining the critical features necessary for accurate predictions.

The introduction of additional NS observables, e.g. tidal deformabilities in the input space is left for future work. As more GW observations from binary NS are expected in a near future, a model that includes tidal deformabilities in its conditioning input vector would be a natural next step. Furthermore, the present model is quite flexible enough to make inference on any other thermodynamics quantity of NS matter, e.g., sound velocity. Lastly, a comparison with conditional generative adversarial NN (GANs) \cite{goodfellow2014generative} would constitute an interesting exploratory work. 

\begin{acknowledgments}
This work was supported by FCT - Fundação para a Ciência e Tecnologia, I.P. through the projects UIDB/04564/2020 and UIDP/04564/2020, with DOI identifiers 10.54499/UIDB/04564/2020 and 10.54499/UIDP/04564/2020, respectively, and the project 2022.06460.PTDC with the associated DOI identifier
10.54499/2022.06460.PTDC, as well as the Polish National Science Centre grant no. 2021/43/B/ST9/01714 and the Italian Ministry of Education, University and Research PRIN project no. 202275HT58.
\end{acknowledgments}

%\bibliography{biblio}

\providecommand{\noopsort}[1]{}\providecommand{\singleletter}[1]{#1}%

\end{document}